\documentclass{ieeetmlcn}

\makeatletter
\def\seclogo{}
\makeatother

\usepackage{algorithm}
\usepackage{algorithmic}
\usepackage{graphicx}
\usepackage{amsmath}
\usepackage{amssymb}
\usepackage{booktabs}
\usepackage{subfigure}      
\usepackage{url}
\usepackage{array,tabularx}
\usepackage[utf8]{inputenc}
\usepackage[T1]{fontenc}

\begin{document}

\receiveddate{26 March, 2025}
\reviseddate{26 August, 2025}
\accepteddate{XX Month, XXXX}
\publisheddate{XX Month, XXXX}
\currentdate{\today}
\doiinfo{TMLCN.XXXX.XXXXXXX}

\title{Towards Intelligent Spectrum Management: Spectrum Demand Estimation Using\\ Graph Neural Networks}

\author{
Mohamad Alkadamani\textsuperscript{\textbf{1,2}},
Amir Ghasemi\textsuperscript{\textbf{1}},
and Halim Yanikomeroglu\textsuperscript{\textbf{2}}
}

\affil{Communications Research Centre Canada, Ottawa, ON, Canada}
\affil{Carleton University, Ottawa, ON, Canada}

\corresp{Corresponding author: Mohamad Alkadamani (email: mohamad.alkadamani@ised-isde.gc.ca).}

\begin{abstract}
The growing demand for wireless connectivity, combined with limited spectrum resources, calls for more efficient spectrum management. Spectrum sharing is a promising approach; however, regulators need accurate methods to characterize demand dynamics and guide allocation decisions. This paper builds and validates a spectrum demand proxy from public deployment records and uses a graph attention network in a hierarchical, multi-resolution setup (HR-GAT) to estimate spectrum demand at fine spatial scales. The model captures both neighborhood effects and cross-scale patterns, reducing spatial autocorrelation and improving generalization. Evaluated across five Canadian cities and against eight competitive baselines, HR-GAT reduces median RMSE by roughly 21\% relative to the best alternative and lowers residual spatial bias. The resulting demand maps are regulator-accessible and support spectrum sharing and spectrum allocation in wireless networks.
\end{abstract}


\maketitle

\section{Introduction}

\IEEEPARstart{T}{he} growing demand for mobile broadband, the rise of smart cities, and the move toward 6G have intensified competition for radio spectrum~\cite{matinmikko2020spectrum}. Mobile data traffic is projected to nearly triple by 2030 \cite{ericsson2019mobility}, reinforcing the need for efficient spectrum management. Dynamic spectrum access and sharing are key strategies~\cite{article12,baldini2020ml}, however, these approaches require fine-grained, spatially localized estimates to guide allocation and sharing decisions.

Conventional spectrum estimation relies on theoretical models or broad indicators (e.g., population density), can miss variations arising from contextual, behavioral, and socio-economic dynamics~\cite{telecom2020ai}. Moreover, proprietary traffic measurements are rarely available to regulators, limiting direct analysis of real-world usage and hindering comparability across regions. These gaps motivate data-driven approaches that leverage open datasets to identify congestion and underutilization, thereby supporting adaptive spectrum allocation~\cite{aygul2025ml}.

Recent advances in AI/ML enable learning at scale from geospatial data~\cite{sabir2024ai}. Graph neural networks (GNNs) are well-suited because they encode neighborhood structure and can mitigate spatial autocorrelation that limits generalization in conventional ML~\cite{jing}. Explicit spatial modeling makes GNNs a practical basis for demand estimation in regulatory contexts. 

This work develops a three-stage pipeline. First, a tile-level \emph{deployment-based indicator} is constructed from public records (deployed bandwidth) and statistically validated against busy-hour downlink throughput from a mobile network operator (MNO). The validated deployment-based indicator is then produced across all study geographies and used as a regulator-accessible supervisory \emph{proxy} for spectrum demand. Second, heterogeneous open datasets are co-registered to a multi-resolution tiled grid and used to build a hierarchical graph that exposes both local neighborhoods and cross-scale structure. Third, a hierarchical, multi-resolution graph attention model (HR-GAT) is trained on this graph; spatially disjoint evaluation protocols curb geospatial information leakage and assess transferability across cities. The contributions are threefold:

\begin{itemize}
\item \textbf{Validated public proxy for spectrum demand:} A regulator-accessible, tile-level proxy derived from public deployed bandwidth and statistically validated against MNO busy-hour throughput.
\item \textbf{Hierarchical, multi-resolution graph attention:} A cross-scale graph design with node-adaptive fusion that captures intra- and inter-scale dependencies, improving generalization under strong geospatial correlation.
\item \textbf{Demand drivers for policy:} A policy-facing attribution analysis identifies the most influential geospatial, demographic, mobility, and economic factors associated with spectrum demand.
\end{itemize}

The remainder of the paper is organized as follows: Section~\ref{sec:relatedwork} reviews related work; Section~\ref{sec:problemformulation} formulates the problem; Section~\ref{sec:methodology} details the methodology; Section~\ref{sec:proxyvalidation} presents the proxy development and validation; Section~\ref{sec:ai_modeling} describes the modeling framework (HR-GAT); Section~\ref{sec:results} reports results and policy-facing analysis; and Section~\ref{sec:conclusion} concludes the paper.

\section{Related Work}\label{sec:relatedwork} 

\subsection{Traditional Approaches}
Spectrum demand estimation has traditionally relied on theoretical models and broad demographic indicators. The International Telecommunication Union (ITU) methodology~\cite{itu2020} uses market data, service usage, and projected traffic distributions; the Federal Communications Commission (FCC) applies empirical models~\cite{fcc2020} incorporating traffic growth, infrastructure expansion, and spectral efficiency. The GSMA mid-band report~\cite{gsma2025} estimates demand for 36 cities via data-rate assumptions, peak concurrency, and offloading. Other works employ statistical trends and capacity-based methods tied to site utilization~\cite{wibisono2015}, as well as queuing and traffic models that often assume uniform demand, missing real spatial variability~\cite{irnich2004spectrum}. A recent time-series approach projects spectrum efficiency based on average per-site metrics for several countries~\cite{SHAYEA20228051}. These methods provide useful baselines but generally lack fine spatial granularity and often depend on proprietary inputs, limiting their use for high-resolution, regulator-accessible estimation.

\subsection{ML Approaches}
Machine learning methods leverage geospatial and infrastructure datasets to estimate demand when direct traffic is unavailable. A tower-count proxy with geospatial factors was introduced in~\cite{ParekhFNWF2023}, but with limited scope and no validation. A follow-up emphasized interpretability~\cite{ParekhPimrc2023} but did not test scalability across diverse urban settings. A proxy derived from licensed holdings was validated against real traffic in two Canadian cities~\cite{Alkadamani2024}, improving confidence yet leaving cross-region generalization open. Deep learning has also been explored. DeepSpectrum~\cite{10757733} applies CNNs and multi-task learning to open data, reducing the need for feature engineering; however, CNNs lack explicit neighborhood modeling on irregular geographies, limiting their ability to capture spatial dependencies.

\subsection{Graph Neural Networks (GNNs) for Spatial Modeling}
GNNs represent geography as graphs and capture adjacency-based dependencies, which is well-suited to spatial regression. A spatial regression GCN bridged deep learning and spatial analytics in~\cite{article11}, and spatial GNNs have been used for demand forecasting in marketplace settings~\cite{app14166989}. For cellular networks, recent work models spatial and temporal fluctuations for traffic prediction~\cite{hou2024graphconstructionflexiblenodes}; a survey highlights deep models’ ability to capture spatio-temporal structure~\cite{Chong}. While single-resolution graphs can suffice when targets vary smoothly at that scale, regulatory demand mapping at fine tiles exhibits strong cross-scale coupling (e.g., supply planned at coarser units spilling into smaller tiles). In such cases, adjacent signals at one resolution can be misleading without multi-resolution context. In contrast, the present study applies graph attention in a hierarchical, multi-resolution, regulator-aligned setting supervised by a validated public proxy.

\section{Problem Formulation}\label{sec:problemformulation}

This study targets long-term spectrum demand estimation at fine spatial granularity to inform licensing and regulatory planning. The focus is on persistent patterns that drive multi-year decisions rather than short-term traffic fluctuations.

Persistent capacity pressure in cellular systems manifests during the busy hour and is routinely used for capacity provisioning. Aggregating busy-hour throughput at suitable horizons reduces stochastic fluctuations and reveals persistent load levels. At the operator level, the principal lever to meet such demand is the amount of actively deployed spectrum in a locality; therefore, deployed bandwidth should increase monotonically with sustained busy-hour load. This relationship is validated by regressing tile-level busy-hour traffic on a public, tile-level deployment measure. Once validated, that measure serves as a regulator-accessible demand proxy across all study geographies. The proxy reflects active deployment rather than nominal holdings and aligns with regulatory spatial resolution.

Formally, let the study area be partitioned into grid tiles \(g\). For each tile, the goal is to estimate a demand proxy \(y_g\) from non-technical geospatial covariates \(\mathbf{x}_g\) (e.g., population, land use, mobility, economic indicators) with spatial context:
\[
y_g = f\!\big(\mathbf{x}_g,\, \mathcal{N}(g),\, s_g\big) + \epsilon,
\]
where \(\mathcal{N}(g)\) denotes the spatial neighborhood of \(g\), \(s_g\) indexes the spatial scale (resolution) at which \(g\) is represented, and \(\epsilon\) is residual error. Estimation is performed on a multi-resolution grid with three nested spatial scales to expose cross-scale structure.

\section{Methodology}\label{sec:methodology}

\begin{figure*}
\centering
\includegraphics[scale=0.55]{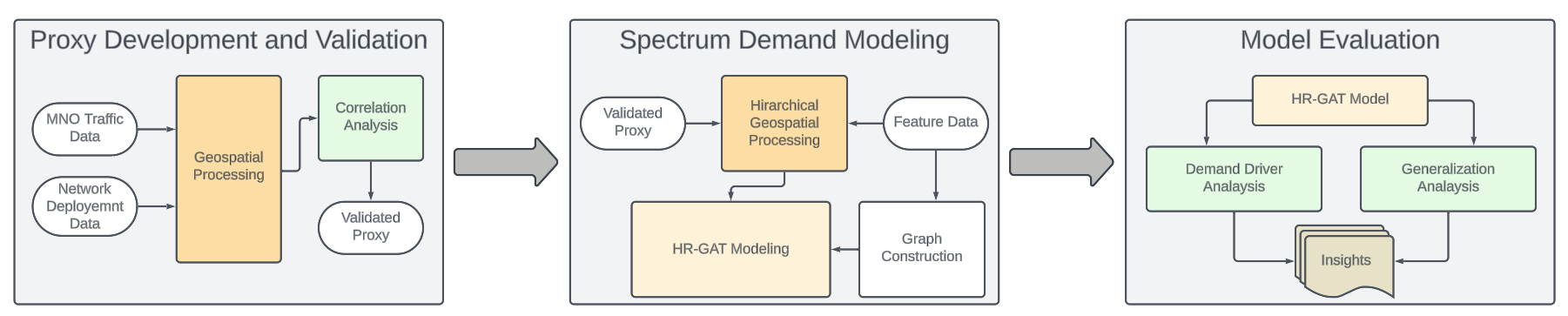} 
\caption{Overview of the proposed methodology.}
\label{fig:spectrum_regulator_actions}
\end{figure*}

The methodology comprises three stages, as shown in Fig.~\ref{fig:spectrum_regulator_actions}.
\begin{enumerate}
    \item Proxy Development and Validation: Public deployment records are distilled into a tile-level signal that reflects actively deployed spectrum bandwidth. Busy-hour MNO traffic is aligned to the same grid and used to validate that signal as a regulator-accessible proxy for sustained demand. Validation is performed via standard correlation and regression analysis; once validated, the proxy is produced across all study geographies and used as the supervisory target \(y_g\) for subsequent modeling.
    \item Spectrum Demand Modeling: Open geospatial datasets are harmonized to the grid and aligned with the validated proxy. Features from multiple sources are spatially aligned and aggregated to capture both local context and broader regional patterns. These inputs are organized into a hierarchical, multi-resolution graph that links tiles across scales, enabling representation of neighborhood effects and cross-scale dependencies. A hierarchical graph attention network (HR-GAT) is trained on this structure to learn relationships between geospatial features and the proxy signal, providing localized spectrum demand estimates.
    \item Model Evaluation and Interpretation: The trained HR-GAT model is evaluated using spatially disjoint validation schemes to assess its generalization across different geographies. Model outputs are further analyzed to identify the main drivers of spectrum demand and to extract insights relevant to regulatory planning and adaptive licensing.
\end{enumerate}

\section{Proxy Development and Validation}
\label{sec:proxyvalidation}

This section constructs a regulator-accessible target \(y_g\) at the tile level by validating a public deployment signal against operator busy-hour traffic. The procedure consists of (i) mapping mobile network operator (MNO) traffic to tiles, (ii) constructing a deployment-based proxy on the same grid, and (iii) validating the proxy via regression.

\subsection{Mapping Operator Traffic to Tiles}

\begin{figure}
    \centering
    \includegraphics[width=0.95\columnwidth]{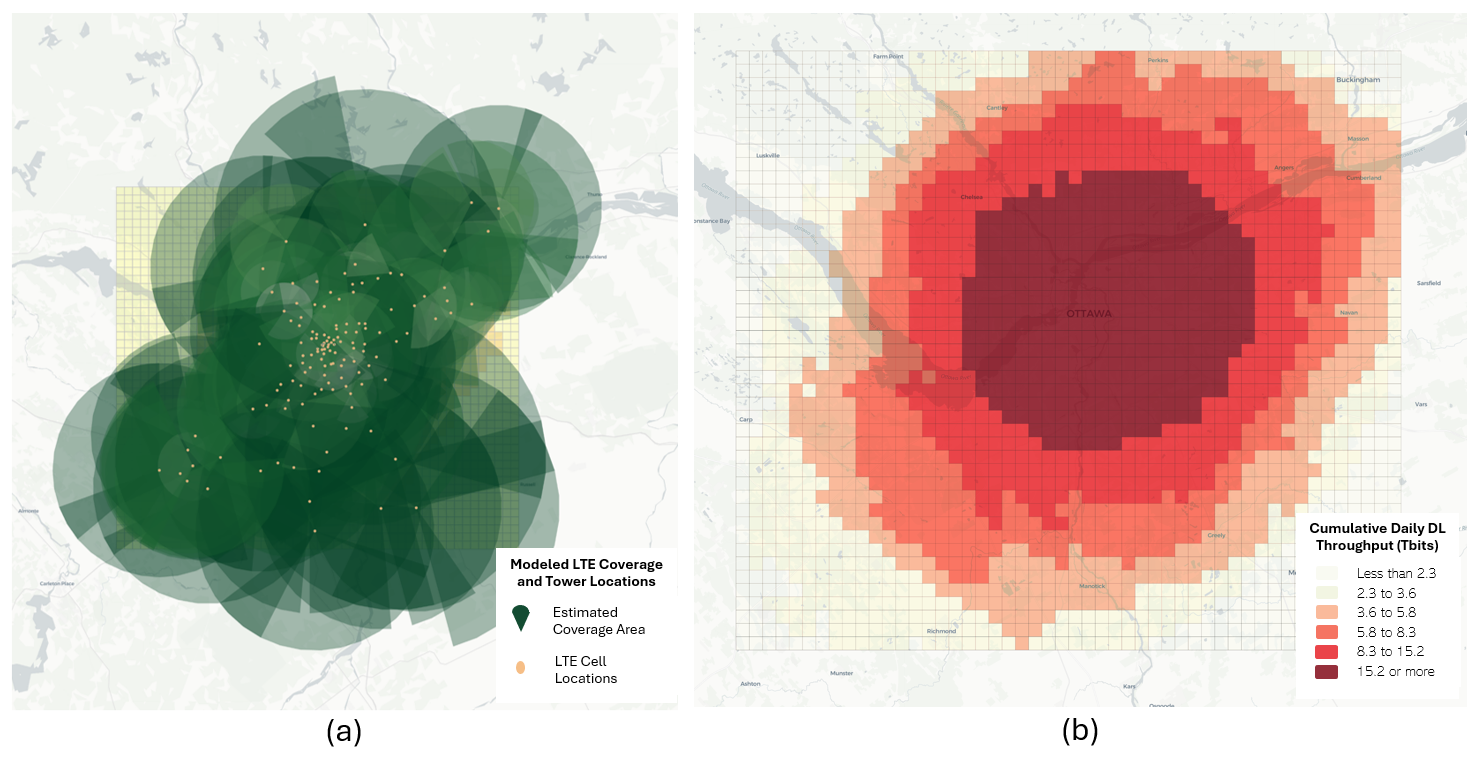}
    \caption{Geospatial processing for proxy validation: (a) LTE cell locations and estimated coverage areas; (b) aggregated busy-hour throughput per grid tile.}
    \label{fig:geo_processing}
\end{figure}

Let \( \mathcal{I} \) be the set of LTE cells, \( \mathcal{G} \) the set of tiles (zoom 15), \( \mathcal{D} \) the set of days, and \( \mathcal{H} \) the set of hours. For each cell \( i \in \mathcal{I} \), let \( t_{i,h,d} \) denote downlink throughput at hour \( h \in \mathcal{H} \) on day \( d \in \mathcal{D} \). The average busy-hour throughput per cell is
\[
\bar{T}_i \;=\; \frac{1}{|\mathcal{D}|} \sum_{d \in \mathcal{D}} \max_{h \in \mathcal{H}} \, t_{i,h,d}.
\]
Coverage is estimated using the e-Hata model~\cite{1622772}. Let \( c_{i g} \in [0,1] \) denote the fraction of tile \( g \in \mathcal{G} \) covered by cell \( i \) (area overlap). The load of cell \( i \) is distributed proportionally over its covered tiles:
\[
\pi_{i g} \;=\; 
\begin{cases}
\displaystyle \frac{c_{i g}}{\sum_{g' \in \mathcal{G}} c_{i g'}} , & \sum_{g'} c_{i g'} > 0, \\[1.2ex]
0, & \text{otherwise}.
\end{cases}
\]
The tile-level traffic target is then
\[
T_g \;=\; \sum_{i \in \mathcal{I}} \bar{T}_i \, \pi_{i g}, 
\qquad g \in \mathcal{G}.
\]
Tiles with LTE coverage below 50\% (\(\sum_i c_{i g} < 0.5\)) are excluded to ensure reliability. Figure~\ref{fig:geo_processing} illustrates: (a) LTE sites (yellow) and modeled sector coverage (dark green) over the zoom-15 tile; (b) the resulting tile heatmap of cumulative daily busy-hour throughput (darker red = higher demand).

\subsection{Constructing the Deployment-Based Proxy}

Let \( \mathcal{S} \) be the set of public deployment records (sites/sectors). For site \( s \in \mathcal{S} \), let \( \mathrm{BW}_s \) denote allocated bandwidth and, where available, \(P_s\) the average EIRP. Because sector azimuths and beamwidths were not provided in the public dataset at the time of the study, coverage is approximated with an omnidirectional pattern to obtain \( \tilde{c}_{s g} \in [0,1] \), the fraction of tile \( g \) covered by site \( s \).

A per-site weighting factor \( \phi_s \) incorporates transmit-power information when present:
\[
\phi_s \;=\; 
\begin{cases}
P_s / \bigl(\frac{1}{|\mathcal{S}|}\sum_{s'} P_{s'}\bigr), & \text{if } P_s \text{ is available},\\
1, & \text{otherwise}.
\end{cases}
\]
Allocate each site's bandwidth proportionally over its covered tiles:
\[
\omega_{s g} \;=\; 
\begin{cases}
\displaystyle \frac{\tilde{c}_{s g}\,\phi_s}{\sum_{g' \in \mathcal{G}} \tilde{c}_{s g'}}, & \sum_{g'} \tilde{c}_{s g'} > 0, \\[1.2ex]
0, & \text{otherwise}.
\end{cases}
\]
The tile-level deployment proxy (used later as \(y_g\)) is
\[
\mathrm{BW}_g \;=\; \sum_{s \in \mathcal{S}} \mathrm{BW}_s \, \omega_{s g}, 
\qquad g \in \mathcal{G}.
\]

\subsection{Validation via Regression}

The proxy is validated against the operator traffic target using ordinary least squares (OLS) on the set of reliable tiles \( \mathcal{G}_{\mathrm{val}} \subseteq \mathcal{G} \):
\[
T_g \;=\; \beta_0 \;+\; \beta_1 \,\mathrm{BW}_g \;+\; \varepsilon_g, 
\qquad g \in \mathcal{G}_{\mathrm{val}}.
\]
Model fit is assessed by the coefficient of determination \(R^2\); overall significance is evaluated via the F-statistic and \(p\)-value. The resulting statistics (reported in Section~\ref{sec:results}) confirm that the deployment-based signal tracks busy-hour demand and can serve as a regulator-accessible supervisory target across cities. For subsequent modeling, the validated deployment-based proxy $\mathrm{BW}_g$ is denoted $y_g$ and used as the supervisory target at tile $g$.

\section{Spectrum Demand Modeling}\label{sec:ai_modeling}

\begin{figure*}[h]
\centering
\includegraphics[scale=0.5]{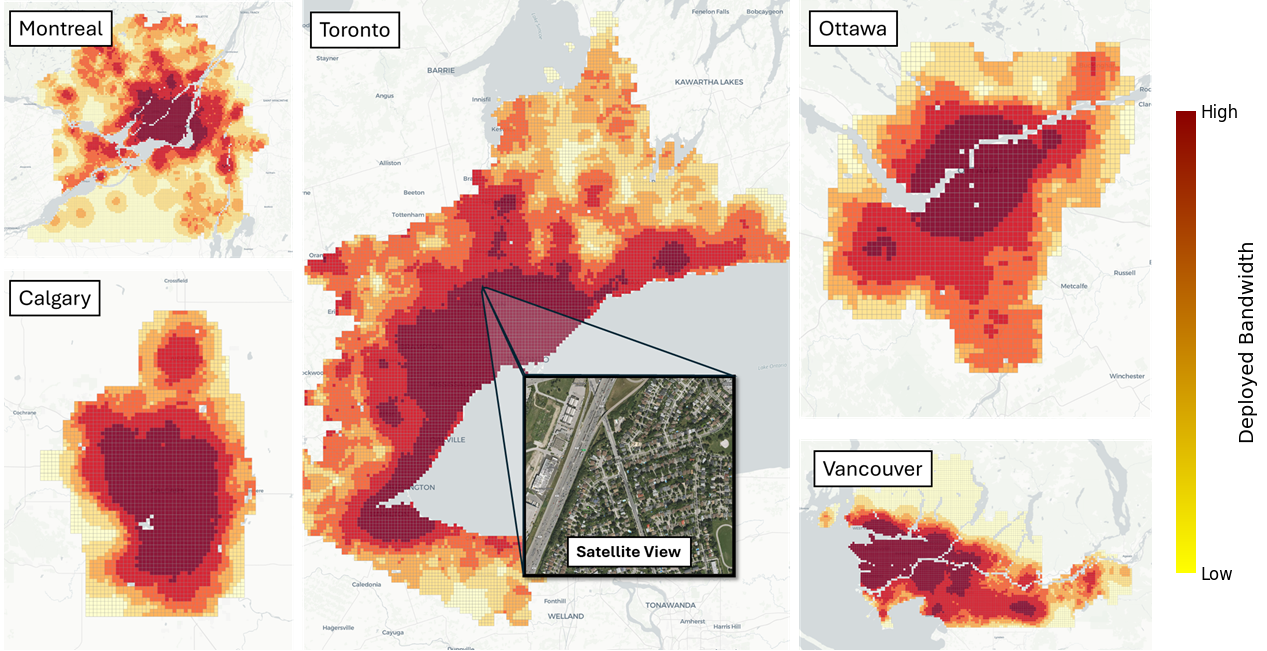} 
\caption{Deployed bandwidth heatmaps at zoom level 15 for Calgary, Montreal, GTA, Vancouver, and Ottawa.}
\label{fig:deployed_bw_heatmap}
\end{figure*}

Following the proxy development, the AI-based modeling framework is applied across five major urban areas in Canada: Vancouver, Calgary, Greater Toronto Area (GTA), Ottawa, and Montreal. These cities were selected to ensure generalizability, as they exhibit distinct demographic, economic, and mobility characteristics~\cite{StatisticsCanada2023}. The objective is to develop an interpretable AI model that estimates spectrum demand across different spatial resolutions.

As detailed in Section~\ref{sec:problemformulation}, the modeling framework adopts a hierarchical, multi-resolution grid structure using Bing Maps tiles. This design enables model training and inference at varying spatial granularities. In total, 50,679 grid tiles are used across the five cities, capturing both dense urban cores and surrounding suburban regions.

To illustrate the spatial distribution of the deployed bandwidth, Fig.~\ref{fig:deployed_bw_heatmap} presents heatmaps for the five study areas at the highest spatial resolution (zoom level 15). The maps depict spatial variation in estimated bandwidth deployment, highlighting regions of high and low spectrum supply. The variation across cities underscores differences in network infrastructure deployment, urban density, and economic activity. Notably, the GTA, Vancouver, and Montreal exhibit extensive high-demand regions, whereas Calgary and Ottawa display more localized concentrations of high-deployment areas, particularly around their urban cores. This spatial characterization provides essential insights into regional disparities in spectrum allocation and network provisioning.

Each zoom level represents a different spatial granularity at which demand drivers (features) and spectrum demand proxy (target variable) are mapped. Given that different data sources exhibit varying spatial distributions, multi-resolution representation allows the model to optimize feature aggregation while mitigating information loss.

\begin{figure}
    \centering
    \includegraphics[width=0.95\columnwidth]{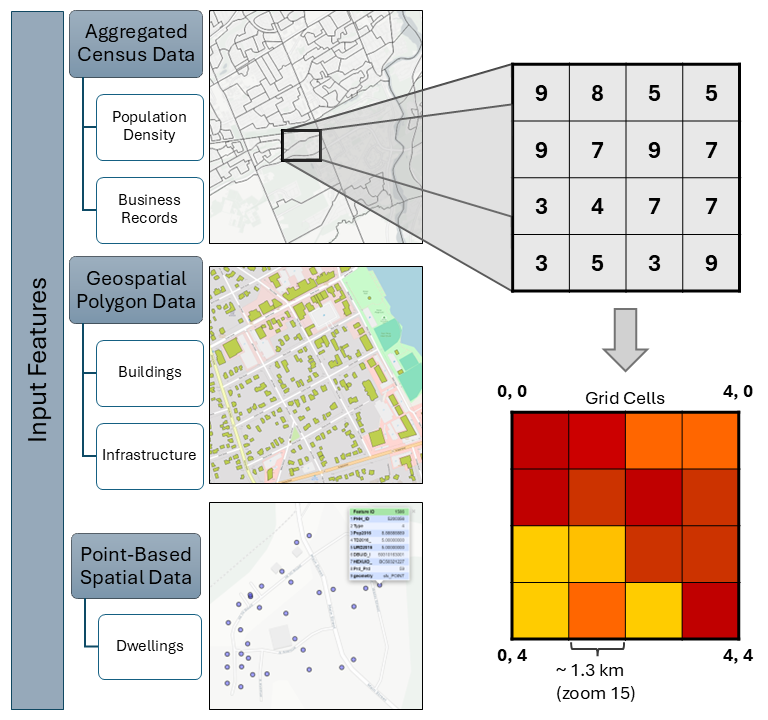}
    \caption{Feature processing pipeline for geospatial demand modeling.}
    \label{fig:feature_processing}
\end{figure}

\subsection{Feature Geospatial Processing}

The framework ingests heterogeneous geospatial datasets that act as potential demand drivers. All inputs are projected onto a uniform Bing-tile grid at zoom levels 13, 14, and 15 to align spatial support and limit aggregation loss.

\textit{Input types:} Inputs fall into three categories: (1) point-based (e.g., households, POIs) aggregated by summing points within each tile; (2) polygon-based with metrics (e.g., census areas with counts) distributed to overlapping tiles using area-weighted interpolation; and (3) polygon-based without metrics (e.g., roads, building footprints) from which tile-level features are derived.

\textit{Data quality:} For point-based inputs, invalid coordinates are removed, duplicates within a tile are deduplicated, empty tiles are set to zero, and isolated gaps are smoothed with a local neighborhood mean. For polygon-based with metrics, attributes are checked for plausibility; missing values at a lower administrative unit are imputed from the enclosing higher-level unit (e.g., lower/upper census units such as dissemination areas (DA) and aggregate dissemination areas (ADA) in Canada), then any remaining gaps are smoothed using adjacent-tile means after interpolation. For polygon-based without metrics, geometries are validated and repaired, lightly simplified or snapped to the grid if needed, tile metrics are computed, empty intersections are set to zero, and geometry-induced outliers are clipped. Figure~\ref{fig:feature_processing} overviews the end-to-end processing and harmonization of inputs onto the multi-resolution grid. A summary of all processed features appears in Table~\ref{tab:features_summary}.

\begin{table}[h]
\centering
\caption{Summary of features processed for demand modeling}
\label{tab:features_summary}
\renewcommand{\arraystretch}{1.2}
\begin{tabularx}{\columnwidth}{|>{\raggedright\arraybackslash}p{3.0cm}|>{\raggedright\arraybackslash}X|}
\hline
\textbf{Feature Category} & \textbf{Description (per grid tile)} \\ \hline
\textbf{Population \& Demographics} & Residents by age, income, and worker type \\ 
\textbf{Daytime Population} & Workers/visitors present during working hours (non-residents included) \\
\textbf{Households} & Number of dwelling units \\ 
\textbf{Mobility (trip distance)} & Counts/shares traveling in distance bands (e.g., 0–3, 3–7, 7–10, 10–15 km) \\ 
\textbf{Economic Activity} & Business counts by sector (e.g., NAICS) and estimated employees \\ 
\textbf{Built Environment (Buildings)} & Building count, total built-up area, and building density \\ 
\textbf{Points of Interest (POIs)} & Counts of POIs by category (e.g., retail, education, health, recreation) \\ 
\textbf{Road Network} & Total road length and segment count \\ 
\textbf{Transit Accessibility} & Number of bus stops and rail/transit stations \\ 
\textbf{Nighttime Lights (NTL)} & Mean nighttime light intensity derived from NASA satellite imagery, used as a proxy for economic and commercial activity \\
 \hline
\end{tabularx}
\end{table}

\subsection{Graph Construction for Multi-Resolution Representation}

The geospatial layout of the study area is structured as a hierarchical graph \( G = (V, E) \), where each grid tile serves as a node and spatial relationships define the edges. The multi-resolution representation is based on Bing tiles at the three zoom levels, where:
\begin{itemize}
    \item Nodes (\( V \)): Represent grid tiles at different zoom levels, each associated with a feature vector.
    \item Edges (\( E \)): Capture spatial adjacency and hierarchical relationships between grid tiles.
\end{itemize}

Each node \( v_i \) is associated with a feature vector \( \mathbf{x}_i \in \mathbb{R}^{d} \) containing geospatial attributes. The hierarchical graph connectivity consists of two types of edges:

\begin{itemize}
    \item Intra-Zoom Edges \( (i, j) \in E_{\text{intra}} \): connect nodes within the same zoom level based on spatial adjacency.
    \item Inter-Zoom Edges \( (i, j) \in E_{\text{inter}} \): connect corresponding nodes across different zoom levels to encode hierarchical dependencies.
\end{itemize}

\begin{figure*}[!t]
    \centering
    \subfigure[Illustration of the HR-GAT hierarchical graph construction. (Left) Geospatial organization of grid tiles across multiple zoom levels. (Right) The corresponding network representation.]{
        \includegraphics[scale=0.4]{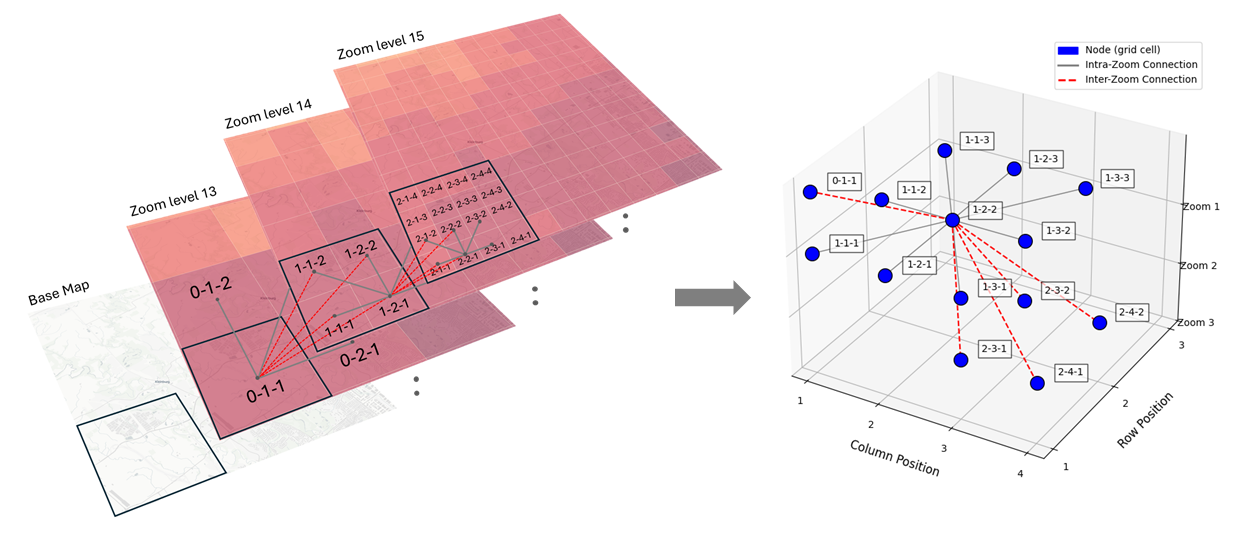}
        \label{fig:hrgat_graph1}
    }
    \hfill
    \subfigure[HR-GAT Architecture for Multi-Zoom Spectrum Demand Estimation.]{
        \includegraphics[scale=0.4]{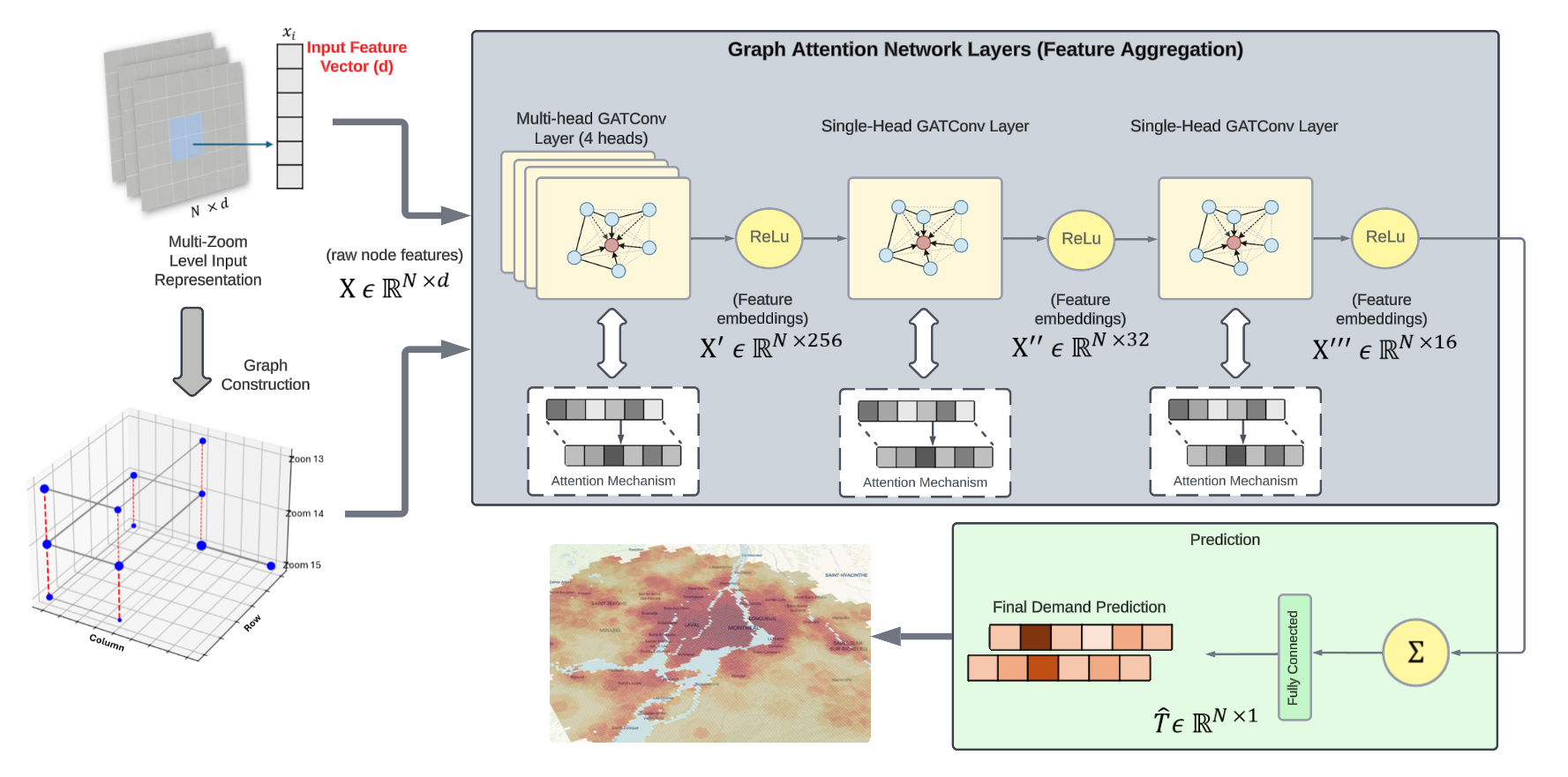}
        \label{fig:hrgat_graph2}
    }
    \caption{Visualization of HR-GAT. (a) Hierarchical graph construction across multiple zoom levels. (b) Model architecture for spectrum demand estimation.}
    \label{fig:hrgat_combined}
\end{figure*}

To formulate the connectivity, adjacency matrices were defined as follows:
\begin{equation}
    A_{\text{intra}} = \{ w_{ij} \ | \ (i, j) \in E_{\text{intra}} \},
\end{equation}
\noindent where \( w_{ij} \) is the edge weight, computed using a Gaussian kernel:
\begin{equation}
    w_{ij} = \exp\left(-\frac{\| s_i - s_j \|^2}{\sigma^2}\right),
\end{equation}
where \( s_i \) and \( s_j \) are the spatial coordinates of tiles \( i \) and \( j \), and \( \sigma \) controls the edge strength.

\begin{equation}
    A_{\text{inter}} = \{ w_{ij} \ | \ (i, j) \in E_{\text{inter}} \}, 
\end{equation}
where \(w_{ij} = 1\) is used for parent-child links.
The graph construction is hierarchical, ensuring that lower-resolution nodes (e.g., zoom 13) aggregate information from finer-resolution nodes (e.g., zoom 15). The final hierarchical graph structure is illustrated in Fig.~\ref{fig:hrgat_combined}~(a), where gray solid edges indicate intra-zoom connections and red dashed edges represent inter-zoom dependencies.

\subsection{HR-GAT Model Architecture}

HR-GAT applies graph attention to regulatory geographies in a hierarchical, multi-resolution setup, modeling both local (intra-scale) and cross-scale dependencies for spectrum demand estimation. The architecture is composed of the following components:

\begin{itemize}
    \item \textbf{Input layer:} Embeds raw geospatial feature inputs for each grid tile into a higher-dimensional latent space.
    \item \textbf{Graph attention layers:} Apply attention-based message passing within each resolution (zoom level) to capture neighborhood dependencies.
    \item \textbf{Cross-scale fusion (node-adaptive):} Learns how to combine information from multiple resolutions through node-specific attention weights.
    \item \textbf{Output layer:} Predicts the final spectrum demand estimate for each grid tile.
\end{itemize}

The input to HR-GAT is the node feature matrix
\begin{equation}
    X = [x_1, x_2, \dots, x_N] \in \mathbb{R}^{N \times d},
\end{equation}
where \(x_i\) is the feature vector of node \(v_i\), and \(d\) is the feature dimension.

Each graph attention layer (GATConv) follows the standard attention-based message passing formulation:
\begin{equation}
    h_i^{(l+1)} = \sigma \left( \sum_{j \in \mathcal{N}(i)} \alpha_{ij}^{(l)} W^{(l)} h_j^{(l)} \right),
\end{equation}
where:
\begin{itemize}
    \item \(h_i^{(l)}\) is the embedding of node \(i\) at layer \(l\),
    \item \(W^{(l)}\) is a learnable weight matrix,
    \item \(\alpha_{ij}^{(l)}\) is the attention coefficient determining the influence of neighbor \(j\),
    \item \(\mathcal{N}(i)\) is the neighborhood of node \(i\), and
    \item \(\sigma(\cdot)\) is a non-linear LeakyReLU activation.
\end{itemize}

The attention coefficients are computed as:
\begin{equation}
    \alpha_{ij} = 
    \frac{\exp \left( \text{LeakyReLU} \!\left( a^{\top} [W h_i \, || \, W h_j] \right) \right)}
    {\sum_{k \in \mathcal{N}(i)} \exp \left( \text{LeakyReLU} \!\left( a^{\top} [W h_i \, || \, W h_k] \right) \right)},
\end{equation}
where \(a\) is a learnable attention vector, \(W\) is the shared transformation matrix, and \(||\) denotes concatenation.

After intra-scale message passing, embeddings from all zoom levels (\(z \in \{13,14,15\}\)) are combined using learnable, node-specific gates that determine the relative importance of each resolution for each tile:
\begin{equation}
\alpha_i^{(z)} = 
\frac{\exp\{\mathbf{q}^\top \tanh(\mathbf{W}_z h_i^{(z)})\}}
{\sum_{z' \in \{13,14,15\}} \exp\{\mathbf{q}^\top \tanh(\mathbf{W}_{z'} h_i^{(z')})\}},
\end{equation}
\begin{equation}
h_i^{\text{final}} = \sum_{z \in \{13,14,15\}} \alpha_i^{(z)} h_i^{(z)},
\end{equation}
Here:
\begin{itemize}
    \item \(h_i^{(z)}\) is the embedding of node \(i\) at zoom level \(z\),
    \item \(\mathbf{W}_z\) and \(\mathbf{q}\) are learnable projection and query parameters,
    \item \(\alpha_i^{(z)} \ge 0\) with \(\sum_z \alpha_i^{(z)} = 1\), forming a softmax over scales, and
    \item \(\tanh(\cdot)\) provides non-linear gating between resolutions.
\end{itemize}

This formulation makes the fusion \emph{node-adaptive}: the informative scale varies by location. If the weights \(\alpha_i^{(z)}\) were constant across nodes, the formulation would reduce to a fixed, global weighted sum. The Gaussian edge weights \(w_{ij}\) used in graph construction define fixed intra-scale adjacency and are independent of the node-adaptive fusion mechanism, which operates across scales. The overall HR-GAT architecture is illustrated in Fig.~\ref{fig:hrgat_combined}(b), showing hierarchical attention layers and the final demand estimation output.

\subsection{Training and Evaluation}

To ensure robust model generalization, two validation strategies are employed: (1) clustering-based cross-validation (CB-CV) to mitigate spatial autocorrelation, and (2) leave-one-city-out cross-validation (LOCO-CV) to assess generalization across different urban environments. These partitioning strategies are followed by model training, where the HR-GAT model is optimized using a spatially-aware loss function and evaluated using multiple performance metrics.
\subsubsection{CB-CV}
\begin{figure}[h!]
    \centering
    \includegraphics[width=0.9\columnwidth]{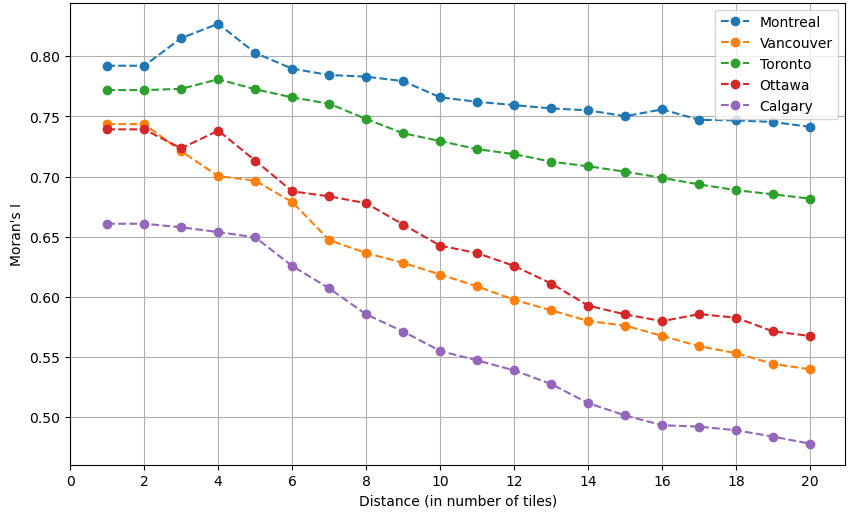}
    \caption{Moran's I analysis of spatial autocorrelation across different cities.}
    \label{fig:moran_i}
\end{figure}

A two-stage clustering scheme is used to curb spatial information leakage and mitigate overfitting. Guided by Moran’s I~\cite{moran}, the analysis in Fig.~\ref{fig:moran_i} shows strong spatial autocorrelation among nearby tiles that decays with distance. This motivates forming compact spatial groups via $k$-nearest neighbors as the basis for cross-validation. The subsequent stage organizes these groups into folds to maintain spatial separation while preserving contextual balance.

The two-stage clustering process proceeds as follows:

\begin{itemize}
    \item \textbf{Stage 1 - spatial proximity clustering (zoom 14):} \ Tiles are grouped on the zoom-14 grid using $k$-nearest neighbors with $k=6$ by geodesic centroid distance, producing compact, contiguous clusters. Edge tiles may form smaller clusters when fewer than six neighbors exist. These clusters are treated as indivisible spatial units in Stage~2 to prevent adjacent tiles from being split across train and validation.

    \item \textbf{Stage 2 - context-aware cluster folding:} Spatial clusters from Stage~1 are grouped into five cross-validation folds that (i) maximize spatial separation between train and validation sets to limit leakage, and (ii) balance land-cover representation. Each cluster receives a dominant land-cover label using ESA WorldCover (10\,m, 2020 v100)~\cite{worldcover2020}, and labeled clusters are packed into five folds to approximate stratified coverage while preserving inter-fold distance. Adjacent tiles are never split across train and validation within a fold, yielding spatially disjoint yet contextually diverse splits that better reflect geospatial generalization.
\end{itemize}

Using this two-stage clustering method, each city’s dataset is divided into five folds where adjacent tiles are never simultaneously included in both training and validation sets, and all major land cover classes are well represented in each fold. This promotes better generalization and more realistic model evaluation, particularly in geospatial contexts where spatial patterns dominate feature distributions.

\subsubsection{LOCO-CV}

To assess generalization to unseen cities, a separate LOCO-CV is performed. In this setting:
\begin{itemize}
    \item The model is trained on four cities and tested on the fifth city, which is entirely unseen.
    \item Ottawa is selected as the test city to evaluate performance on a new urban environment.
\end{itemize}

This approach complements CB-CV by measuring how well the model generalizes to different spatial distributions and urban morphologies.

\subsection{Loss Function}

The HR-GAT model is trained using a combination of Mean Squared Error (MSE) for demand estimation and a spatial regularization term to smooth predictions across connected nodes in the hierarchical graph:

\begin{equation}
    \mathcal{L}_{\text{MSE}} = \frac{1}{N} \sum_{i=1}^{N} \left( y_i - \hat{y}_i \right)^2,
\end{equation}

To prevent excessive local fluctuations in predictions, a spatial regularization term is introduced:

\begin{equation}
    \mathcal{L}_{\text{spatial}} = \frac{1}{|E|} \sum_{(i, j) \in E} \left( \hat{y}_i - \hat{y}_j \right)^2,
\end{equation}

The final objective function is a weighted sum of both terms:

\begin{equation}
    \mathcal{L} = \mathcal{L}_{\text{MSE}} + \lambda \mathcal{L}_{\text{spatial}}.
\end{equation}

\noindent where:
\begin{itemize}
    \item $y_i$ is the ground-truth target for tile (node) $i$,
    \item $\hat{y}_i$ is the model estimate for tile $i$,
    \item $N$ is the number of tiles (nodes),
    \item $E$ is the set of graph edges (intra- and inter-zoom), with $|E|$ its size,
    \item $(i,j)\in E$ are adjacent tiles connected in the hierarchical graph, and
    \item $\lambda>0$ is the weight on the spatial smoothness term.
\end{itemize}

\subsection{Model Evaluation Metrics}

The model is evaluated using the following metrics:

\begin{enumerate}
    \item \textbf{Mean Absolute Error (MAE):} Measures absolute prediction accuracy by averaging the absolute differences between predicted and actual values.

    \item \textbf{Root Mean Squared Error (RMSE):} Measures the standard deviation of the residuals (prediction errors), penalizing larger errors more heavily than MAE.

    \item \textbf{R-Squared (\( R^2 \)) Score:} Evaluates how well the model explains the variance in spectrum demand.
\end{enumerate}

Metrics are aggregated as the median across evaluation tiles and folds.

\section{Performance Evaluation}\label{sec:results}

This section evaluates HR-GAT under two regimes: (i) CB-CV within cities and (ii) a single-city LOCO-CV. The training routine used in both regimes is summarized in Algorithm~\ref{alg:HR-GAT}.

\begin{algorithm}[h!]
\caption{Training procedure for HR-GAT with node-adaptive fusion}
\label{alg:HR-GAT}
\begin{algorithmic}[1]

\REQUIRE Multi-resolution graph $G=(V, A_{\text{intra}}, A_{\text{inter}})$,
node features $X$, target $y$, learning rate $\eta$, loss weight $\lambda$,
and epochs $E$.

\STATE Initialize model parameters
$\{W^{(l)}, a^{(l)}, \mathbf{W}_z, \mathbf{q}\}$.

\FOR{each epoch $e = 1$ to $E$}

\STATE \textbf{(1) Intra-scale attention:}
Update embeddings $h_i^{(z)}$ for each zoom level $z$ using GAT message
passing over $A_{\text{intra}}$.

\STATE \textbf{(2) Cross-scale fusion:}
Compute node-specific fusion weights $\alpha_i^{(z)}$ and obtain
$h_i^{\text{final}}=\sum_{z}\alpha_i^{(z)}h_i^{(z)}$.

\STATE \textbf{(3) Prediction:}
Estimate $\hat{y}_i = f_{\text{out}}(h_i^{\text{final}})$.

\STATE \textbf{(4) Loss:}
Compute $\mathcal{L}$ as defined in Sec.~\ref{sec:ai_modeling}.

\STATE \textbf{(5) Parameter update:}
Apply gradient descent with rate $\eta$ to minimize $\mathcal{L}$.

\ENDFOR

\RETURN trained HR-GAT model.

\end{algorithmic}
\end{algorithm}

\subsection{Proxy Validation Results}

This subsection presents the OLS regression results evaluating the relationship between deployed bandwidth and MNO traffic demand in the study region.

Table~\ref{tab:ols_results} summarizes the results. The coefficient of determination \(R^2 = 0.727\) indicates that deployed bandwidth explains 72.7\% of the variance in traffic demand. The overall regression is significant (\(F = 4477\), \(p < 0.001\)), supporting the reliability of the proxy and its use as the supervisory target in HR-GAT.

\begin{table}[h]
\centering
\caption{ols regression results for proxy validation}
\label{tab:ols_results}
\begin{tabular}{lccc} 
\toprule
\textbf{Proxy Variable} & \textbf{R²} & \textbf{F-Statistic} & \textbf{p-Value} \\ 
\midrule
Deployed Bandwidth ($\mathrm{BW}_g$) & 0.727 & 4477 & $\mathbf{< 0.001}$ \\ 
\bottomrule
\end{tabular}
\end{table}

\subsection{Comparison with Alternative ML Models Using CB-CV}

The proposed HR-GAT model is evaluated against multiple ML baselines, including LightGBM \cite{NIPS2017}, gradient boosting \cite{Friedman2001}, XGBoost \cite{Chen2016}, decision tree \cite{Quinlan1986}, random forest \cite{Breiman2001}, a linear model, a vanilla CNN \cite{OShea2015}, and a plain GAT \cite{Velickovic2017}. The comparison uses three metrics: MAE, RMSE, and \(R^2\), as summarized in Table~\ref{tab:model_comparison}.

Two neural-network baselines are included: a vanilla CNN, which processes each tile independently (no spatial edges), and a plain GAT, which introduces same-resolution adjacency but no cross-resolution transfer.

HR-GAT achieves the best overall performance, with an RMSE of 29.30, an MAE of 10.93, and \(R^2=0.91\). This indicates that explicitly modeling spatial adjacency and hierarchical structure substantially improves accuracy when demand exhibits strong spatial dependencies.

Among the alternatives, the plain GAT ranks second by RMSE (RMSE 37.30; \(R^2=0.83\)). Although graph-based learning helps, the absence of multi-resolution coupling limits its effectiveness relative to HR-GAT. XGBoost and gradient boosting perform competitively (RMSE~\(\approx\)~44; \(R^2=0.86\)) but lack explicit spatial awareness, which can hinder generalization under geographic correlation. LightGBM performs slightly worse (RMSE 55.70). The vanilla CNN underperforms tree-based models (RMSE 69.30; \(R^2=0.63\)), suggesting that added depth without spatial edges does not help. Tree-based baselines (decision tree, random forest) perform poorly (RMSE \(>\) 77), and the linear model is worst (RMSE 89.31; \(R^2=0.43\)). A detailed breakdown is shown in Table~\ref{tab:model_comparison}.

\begin{table}[h]
\centering
\caption{performance comparison of hr-gat vs. alternative ml models}
\label{tab:model_comparison}
\begin{tabular}{|l|c|c|c|}
\hline
\textbf{Model} & \textbf{Median MAE} ↓ & \textbf{Median RMSE} ↓ & \textbf{R²} ↑ \\ 
\hline
Gradient Boosting & 23.44 & 44.27 & 0.86 \\ 
XGBoost & 23.53 & 44.72 & 0.86 \\ 
LightGBM & 30.78 & 55.70 & 0.80 \\ 
Decision Tree & 41.60 & 79.92 & 0.53 \\ 
Random Forest & 45.89 & 77.48 & 0.55 \\ 
Linear Model & 55.50 & 89.31 & 0.43 \\ 
\textbf{Vanilla CNN} & \textbf{44.23} & \textbf{69.30} & \textbf{0.63} \\ 
\textbf{Plain GAT} & \textbf{24.88} & \textbf{37.30} & \textbf{0.83} \\ 
\textbf{HR-GAT (Proposed)} & \textbf{10.93} & \textbf{29.30} & \textbf{0.91} \\ 
\hline
\end{tabular}
\end{table}

\begin{figure}[h] \centering \includegraphics[width=0.95\columnwidth]{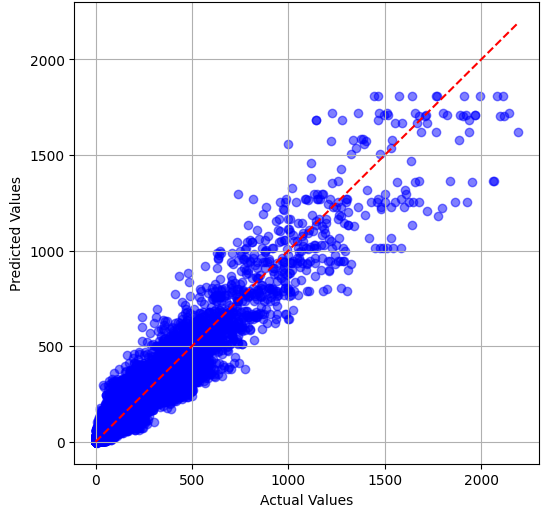} \caption{Scatter plot of actual vs. predicted spectrum demand values for HR-GAT.} \label{fig:scatterplot} \end{figure}

Figure~\ref{fig:scatterplot} presents a scatter plot comparing actual and predicted values using the HR-GAT model. Each point represents a grid tile, with the x-axis indicating actual values and the y-axis representing model estimates. The red dashed line denotes the ideal 1:1 correspondence, where estimates perfectly match actual values. Results indicate that HR-GAT achieves strong predictive performance, with most points closely aligning along the diagonal. However, some deviation is observed at higher demand levels, indicating slight underestimation in high-traffic regions. This suggests that while HR-GAT effectively captures spatial demand patterns, further refinements could improve its handling of extreme values.

\begin{figure}
    \centering
    \includegraphics[width=0.95\columnwidth]{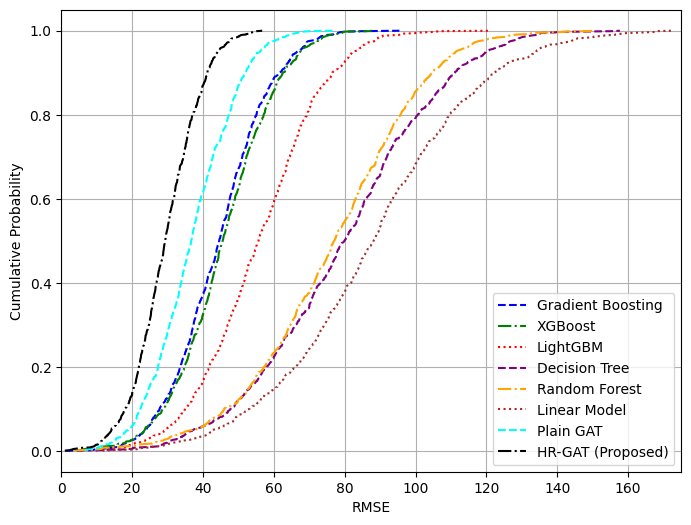}
    \caption{Empirical cumulative distribution function (eCDF) of RMSE values across all models.}
    \label{fig:ecdf_rmse}
\end{figure}

Figure~\ref{fig:ecdf_rmse} presents the empirical cumulative distribution function (eCDF) of RMSE values for all evaluated models. The eCDF provides insight into the distribution of estimation errors, illustrating the proportion of test samples that achieve a given RMSE or lower. The HR-GAT model (black dashed line) demonstrates a significantly steeper curve, indicating that a higher proportion of its estimates fall within lower RMSE values compared to other models. Plain GAT also exhibits strong performance but does not match HR-GAT, reinforcing the impact of hierarchical multi-resolution learning. gradient boosting and XGBoost show comparable error distributions and perform tree-based models such as decision tree and random forest. The linear model performs the worst, with a broad RMSE distribution, highlighting its limited ability to capture complex spatial patterns.

The improved error distribution observed in the eCDF of RMSE highlights HR-GAT’s ability to achieve lower estimation errors compared to other models. However, beyond overall accuracy, it is crucial to assess whether residual errors exhibit spatial autocorrelation, as high spatial dependency in residuals can indicate biases in estimation across different regions.

To further assess the spatial behavior of model errors, Moran’s I is computed on the residuals of HR-GAT, plain GAT, and vanilla CNN \cite{Muller2009}. While Moran’s I was previously used to characterize spatial dependence in the data (Sec.~\ref{sec:ai_modeling} (D)), here it measures residual spatial autocorrelation, indicating whether prediction errors are spatially clustered or random. Lower Moran’s I values correspond to more spatially independent residuals.

The results indicate that HR-GAT achieves the lowest spatial autocorrelation in residuals (Moran’s I = 0.0202), outperforming both plain GAT (0.0253) and vanilla CNN (0.0370). These findings suggest that introducing hierarchical edges and multi-resolution knowledge transfer effectively reduces spatial bias in the predictions. Although the absolute differences in Moran’s I appear small, they represent 20.2\% and 45.4\% reductions relative to the plain GAT and vanilla CNN, respectively. A paired $t$-test across cross-validation folds indicates that the reduction in spatial autocorrelation is statistically significant when compared to both baseline models ($p < 0.05$).

\begin{figure}[h] \centering \includegraphics[width=0.95\columnwidth]{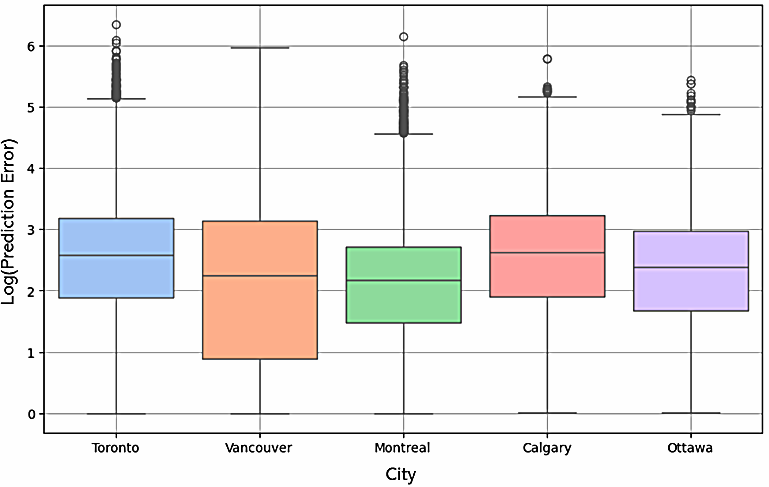} \caption{Log-scaled residual error distribution across cities.} \label{fig:residual_boxplot} \end{figure}

This improvement in generalizability is further supported by the residual error distribution across different cities, as shown in Fig.~\ref{fig:residual_boxplot}. Log-scaled residuals indicate that HR-GAT maintains a consistent error distribution across all five cities, suggesting that the model does not exhibit location-specific biases. While minor variations exist due to local urban characteristics, the comparable interquartile ranges and median errors across cities reinforce the model's ability to generalize effectively beyond the training data.

\subsection{Performance on an Unseen City Using LOCO-CV}
To further evaluate generalization, a LOCO-CV is conducted in which Ottawa is excluded from training and used only for testing. This experiment assesses the ability to generalize to a completely unseen city, simulating a real-world scenario where spectrum regulators may need to transfer knowledge from one set of urban regions to another. Unlike standard cross-validation, this setup provides a more stringent test of model robustness, ensuring that estimation is not biased by spatial dependencies within the training set.

Table~\ref{tab:ottawa_mae} presents the median MAE results for all models when evaluated on Ottawa as an unseen city. HR-GAT achieves the lowest median MAE of 18.74, significantly outperforming all other baselines. The Plain GAT model, which lacks hierarchical resolution transfer, also performs relatively well with a median MAE of 23.30, reinforcing the importance of spatial awareness in predictive modeling. 

Among the tree-based models, Gradient Boosting and XGBoost exhibit the best performance, achieving MAE values of 37.34 and 39.49, respectively. LightGBM performs slightly worse (MAE = 47.20), but still outperforms simpler tree-based approaches. Decision Tree and Random Forest show substantial degradation in performance (MAE $>$ 50), demonstrating their limited ability to generalize across regions with distinct spatial characteristics. Similarly, the Linear Model records the worst MAE of 60.93, reinforcing the limitations of models that fail to capture the complex spatial and nonlinear interactions inherent in spectrum demand estimation. Overall, these findings confirm that HR-GAT’s explicit integration of hierarchical spatial relationships significantly enhances predictive generalization.

\begin{table}[h]
\centering
\caption{median mae comparison for ottawa as the test set}
\label{tab:ottawa_mae}
\begin{tabular}{|l|c|}
\hline
\textbf{Model} & \textbf{Median MAE (Ottawa)} ↓ \\ 
\hline
Gradient Boosting & 37.34 \\ 
XGBoost & 39.49 \\ 
LightGBM & 47.20 \\ 
Decision Tree & 52.47 \\ 
Random Forest & 50.66 \\ 
Linear Model & 60.93 \\ 
\textbf{Vanilla CNN} & \textbf{48.22} \\ 
\textbf{Plain GAT} & \textbf{23.30} \\ 
\textbf{HR-GAT (Proposed)} & \textbf{18.74} \\ 
\hline
\end{tabular}
\end{table}

\subsection{Key Factors Influencing Spectrum Demand}
\begin{figure}[h]
    \centering
    \includegraphics[width=0.95\columnwidth]{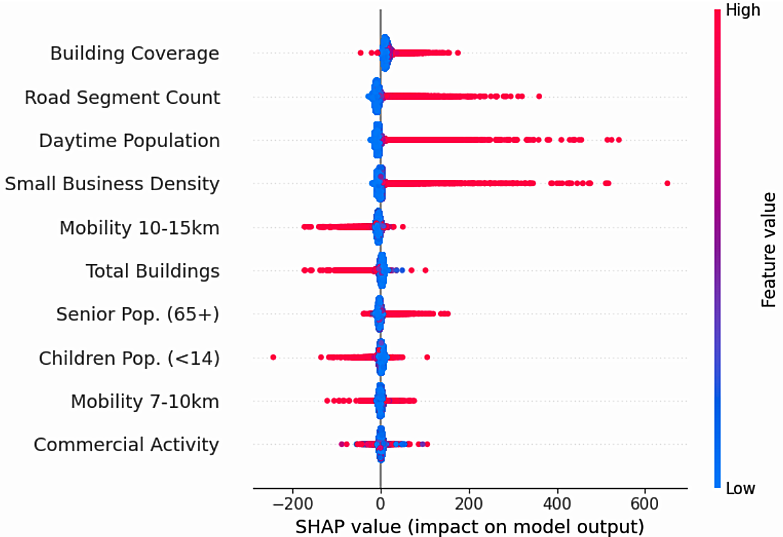}
    \caption{SHAP summary plot illustrating the relative importance of the top 10 features in predicting mobile spectrum demand. Higher absolute SHAP values indicate stronger influence on the model estimates.}
    \label{fig:shap_summary}
\end{figure}

Interpreting estimates from complex ML models, such as HR-GAT, requires techniques that explain the contributions of individual features. Since HR-GAT is a black-box model, the SHapley Additive exPlanations (SHAP) method is used to quantify the impact of each feature on the model’s estimations~\cite{Lundberg2017}. SHAP assigns an importance value to each feature per estimation, enabling an understanding of how geospatial, demographic, and economic factors contribute to spectrum demand.
Figure~\ref{fig:shap_summary} presents the ten most influential features out of 30 input features used in the model.

Urban infrastructure and density play a crucial role in shaping spectrum demand. Features such as building coverage, road segment count, and total number of buildings indicate the extent of urban development, with higher values generally correlating with increased mobile traffic due to dense commercial and residential activities. Similarly, daytime population and small-business density highlight areas of concentrated human presence, reinforcing the that business districts and commercial centers sustain high network loads during peak hours.

Mobility patterns also exhibit a strong influence on spectrum demand. The number of people traveling 7–10 km and 10–15 km capture inter-zonal commuting behavior, where high movement levels correspond to dynamic variations in network traffic, particularly in transit corridors and suburban-urban interfaces.

Demographic factors provide additional insights into demand variations. The presence of senior citizens (65+) and children ($<$14) contributes to household-driven network usage patterns, albeit with a lower overall influence compared to commercial and commuting-related factors.

Finally, commercial activity, approximated through nighttime lights (NTL)~\cite{Li2020}, emerges as a significant determinant, capturing economic hotspots where high levels of business and commercial interactions drive sustained mobile service consumption.

These findings demonstrate that spectrum demand is influenced by a combination of urban infrastructure, population distribution, mobility patterns, and economic activity, emphasizing the need for data-driven spectrum allocation strategies.

\subsection{Implications for Spectrum Policy}

HR-GAT produces regulator-accessible, high-resolution demand maps that support long-term policy choices. Although the present study analyzes a single snapshot and focuses on urban settings, it provides actionable geospatial evidence for licensing and sharing decisions. Temporal dynamics and rural sparsity will be addressed next to strengthen longitudinal and nationwide applicability.

\textbf{Policy-facing implications (current model)}
\begin{itemize}
    \item \emph{Targeted licensing and band planning}—Fine-grained demand maps help identify where additional authorizations or refarming are most warranted and where existing assignments appear sufficient.
    \item \emph{Evidence for spectrum sharing}—Spatial gradients highlight candidate areas for localized or licensed-shared access, informing protection contours, coordination zones, and secondary-use eligibility.
    \item \emph{Timing and phasing of releases}—Cross-city contrasts support phasing of future awards and renewals, aligning auction or assignment timing with demonstrated regional need.
\end{itemize}

\textbf{Scope considerations and planned extensions}
\begin{itemize}
    \item The model does not yet capture diurnal or seasonal variation or multi-year trends; adding temporal conditioning will inform when, as wel as where, sharing or new awards are necessary.
    \item Sparse-node geographies may weaken graph signals; a rural-aware graph is planned to improve estimates in low-density areas and support differentiated rural licensing.
    \item The target is a validated public-data proxy rather than direct traffic; ongoing proxy refinement and cross-source validation will tighten confidence bounds for licensing and sharing decisions.
\end{itemize}

\section{Concluding Remarks}\label{sec:conclusion}
This work develops a regulator-accessible framework for fine-grained spectrum demand estimation by validating a public-data proxy against operator traffic and scaling it across cities, and by applying a hierarchical, multi-resolution graph attention formulation (HR-GAT) tailored to tiled regulatory geographies. The framework integrates open geospatial, demographic, and economic indicators to learn stable spatial patterns that underlie capacity needs.

Empirically, the approach outperforms tree-based models, CNNs, and single-resolution GNNs across five Canadian cities (median RMSE 29.30; \(R^2 = 0.91\)) and yields the lowest residual spatial autocorrelation (Moran’s I), indicating stronger generalization and reduced spatial bias. An ablation with a single-resolution GAT indicates that multi-resolution coupling and node-adaptive fusion materially improve accuracy and robustness.

For policy, the resulting high-resolution demand maps provide reproducible, non-proprietary evidence to (a) target licensing and band refarming where need is most acute, and (b) identify credible candidates for localized or shared access with defensible protection contours. By combining a validated public proxy with a hierarchical application of graph attention, the framework offers a practical, scalable, and transparent basis for spectrum planning and sharing decisions.

\section*{Acknowledgment}
The authors thank Sarah Dumoulin and Colin Brown, Communications Research Centre Canada (CRC), for providing valuable feedback and comments.

\section*{Conflict of Interest}
The authors declare that they have no conflict of interest to disclose.

\bibliographystyle{IEEEtran} 
\bibliography{references}    

\begin{IEEEbiography}[{\includegraphics[width=1in,height=1.25in,clip,keepaspectratio]{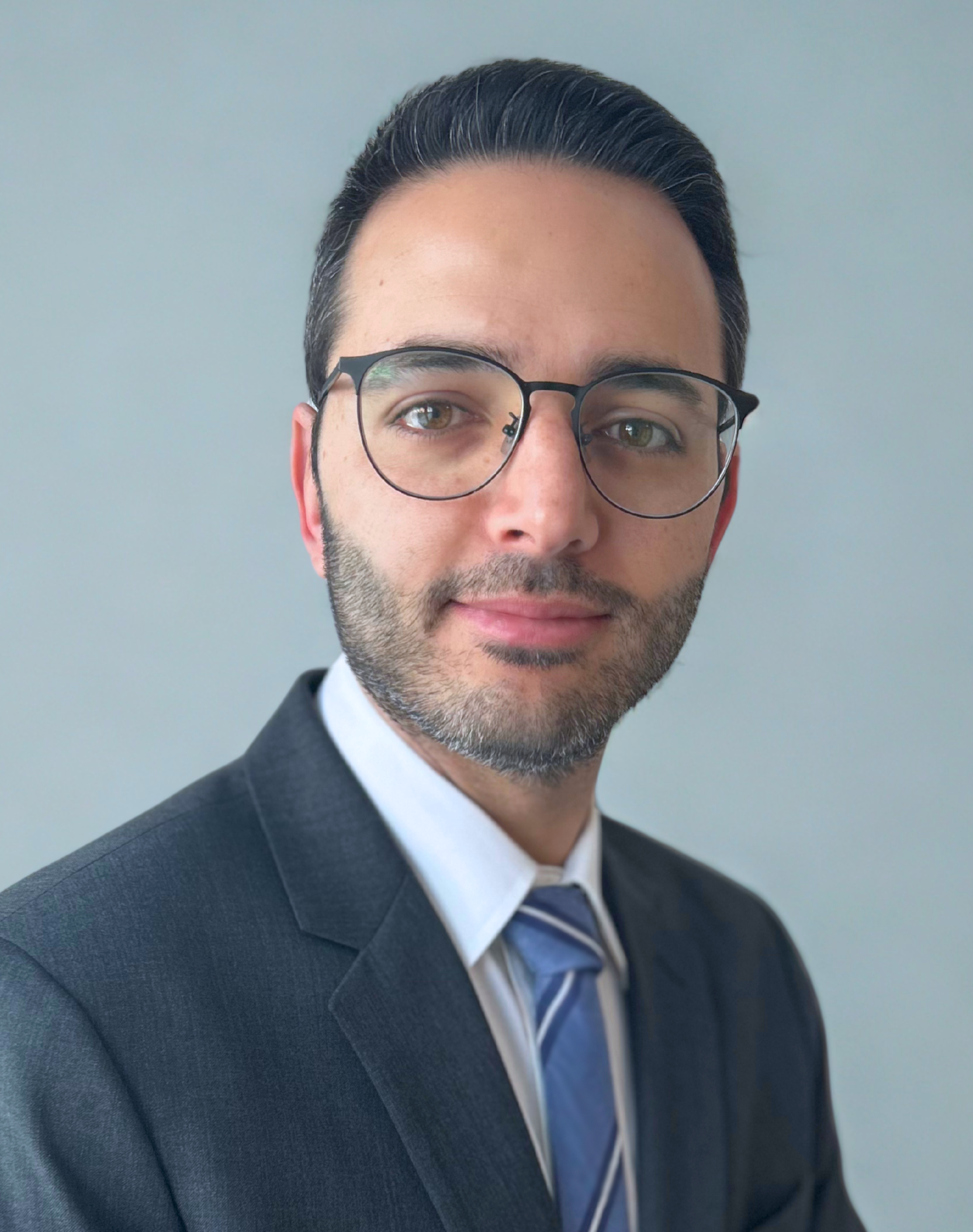}}]{Mohamad Alkadamani}
 received the M.A.Sc. degree in Electrical and Computer Engineering from Carleton University in 2019. He is currently pursuing a Ph.D. in Electrical and Computer Engineering at Carleton University. He is also a Research Engineer with the Data Science Group at the Communications Research Centre Canada, where his work focuses on advancing data-driven methodologies for intelligent spectrum management and optimization. His research interests include machine learning for wireless networks, geospatial data analysis, and spectrum sharing strategies to enhance the efficiency and adaptability of modern wireless communication systems.
\end{IEEEbiography}

\begin{IEEEbiography}[{\includegraphics[width=1in,height=1.25in,clip,keepaspectratio]{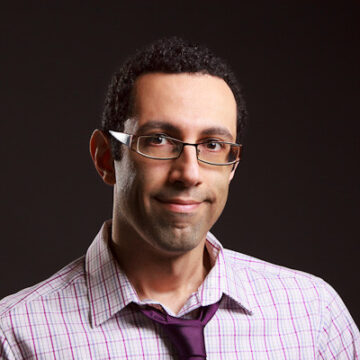}}]{Amir Ghasemi}
received the M.A.Sc. and Ph.D. degrees in electrical and computer engineering from the University of Toronto. He is a Senior Research
Scientist at the Communications Research Centre Canada. He has contributed
to several international wireless communications standards and authored
more than 30 research papers on dynamic spectrum access, cognitive radio,
and wireless coexistence. He has held adjunct professorships with Queen’s
University as well as Ontario Tech University. His current research is focused
on data-driven intelligent wireless networking. In 2014, he received the
Governor General’s “Public Service Award of Excellence for Innovation.”
\end{IEEEbiography}

\begin{IEEEbiography}[{\includegraphics[width=1in,height=1.25in,clip,keepaspectratio]{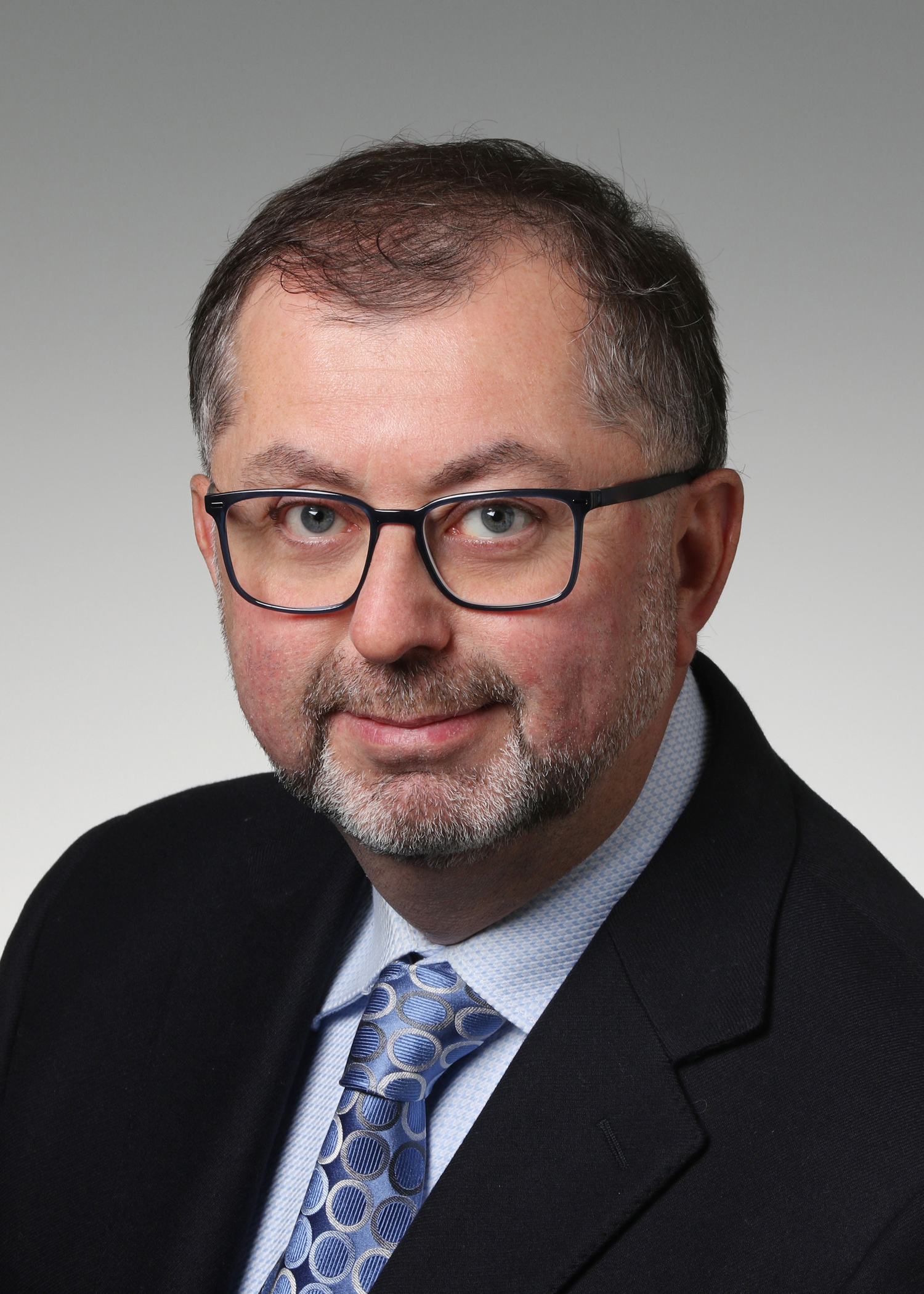}}]{Halim Yanikomeroglu}
(Fellow, IEEE) received
the B.A.Sc. degree in electrical and electronics
engineering from Middle East Technical University, Ankara, Türkiye, in 1990, and the M.A.Sc.
degree in electrical engineering and the Ph.D.
degree in electrical and computer engineering
from the University of Toronto, Toronto, ON,
Canada, in 1992 and 1998, respectively.
Since 1998, he has been with the Department of Systems and Computer Engineering,
Carleton University, Ottawa, ON, Canada, where
he is now a Chancellor’s Professor. He has delivered more than 110 invited
seminars, keynotes, panel talks, and tutorials in the last five years.
He has supervised or hosted over 160 postgraduate researchers in
his laboratory at Carleton University. Dr. Yanikomeroglu’s extensive
collaborative research with industry resulted in 40 granted patents. His
research interests cover many aspects of wireless communications and
networks, with a special emphasis on non-terrestrial networks (NTN) in
the recent years.
Dr. Yanikomeroglu is a Fellow of Engineering Institute of Canada (EIC),
the Canadian Academy of Engineering (CAE), and the Asia–Pacific
Artificial Intelligence Association (AAIA). He is a member of the
IEEE ComSoc Governance Council, IEEE ComSoc GLOBECOM/ICC
Management and Strategy (GIMS), IEEE ComSoc Conference Council,
and IEEE International Symposium on Personal, Indoor, and Mobile
Radio Communications (PIMRC), and IEEE Future Networks World
Forum (FNWF) Steering Committees. He is a Distinguished Speaker
of the IEEE Communications Society and the IEEE Vehicular Technology Society, and an Expert Panelist of the Council of Canadian
Academies (CCA|CAC). He is currently serving as the Chair of the
Steering Committee for the IEEE’s Flagship Wireless Event, Wireless
Communications and Networking Conference (WCNC). He served as
the General Chair and the Technical Program Chair for several IEEE
conferences. He has also served on the editorial boards for various
IEEE periodicals. He received several awards for his research, teaching,
and service, including the IEEE ComSoc Satellite and Space Communications TC Recognition Award in 2023, IEEE ComSoc Fred W.
Ellersick Prize in 2021, IEEE VTS Stuart Meyer Memorial Award in 2020,
and IEEE ComSoc Wireless Communications TC Recognition Award
in 2018. He received best paper awards at the IEEE Competition on
Non-Terrestrial Networks for B5G and 6G in 2022 (Grand Prize), IEEE
ICC 2021, and IEEE WISEE 2021 and 2022.
\end{IEEEbiography}

\end{document}